\documentclass[referee,a4paper,12pt,traditabstract]{swsc} 
\usepackage{graphicx}
\usepackage{amsmath}
\usepackage{txfonts}
\usepackage{subfigure}
\usepackage{epstopdf}
\usepackage{lineno}
\usepackage[authoryear,round]{natbib}
\usepackage[backref]{hyperref}
\usepackage{url}
\hypersetup{colorlinks=true,citecolor=cyan,urlcolor=cyan,linkcolor=blue}
\begin{document}
\title{A Probabilistic Approach to the Drag-Based Model}
\titlerunning{The Probabilistic Drag-Based Model}
\authorrunning{Napoletano et al.}
\author{
	Gianluca Napoletano\inst{1} 
	\and
    Roberta Forte\inst{2}
    \and
    Dario Del Moro\inst{2}
    \and
    Ermanno Pietropaolo\inst{1}
        \and
    Luca Giovannelli\inst{2}
    \and
    Francesco Berrilli\inst{2}
    } 
\institute{
Dipartimento di Scienze Fisiche e Chimiche, Universit\`a degli studi dell'Aquila, Via Vetoio snc, 67100 Coppito (AQ)\\
         \and
Dipartimento di Fisica, Universit\`a degli studi di Roma ''Tor Vergata'', Via della Ricerca Scientifica, 1, 00133 Roma\\
             }
%\date{} % leave empty
%
\abstract
%% context heading (optional). leave {} empty if necessary  
{The forecast of the time of arrival of a coronal mass ejection (CME) to Earth is of critical importance for our high-technology society and for any future manned exploration of the Solar System.
As critical as the forecast accuracy is the knowledge of its precision, i.e. the error associated to the estimate.
We propose a statistical approach for the computation of the time of arrival using the drag-based model by introducing the probability distributions, rather than exact values, as input parameters, thus allowing the evaluation of the uncertainty on the forecast.
We test this approach using a set of CMEs whose transit times are known, and obtain extremely promising results: the average value of the absolute differences between measure and forecast is 9.1h, and half of these residuals are within the estimated errors.\\
These results suggest that this approach deserves further investigation.
We are working to realize a real-time implementation which ingests the outputs of automated CME tracking algorithms as inputs to create a database of events useful for a further validation of the approach.\\   
}
\maketitle
\section{Introduction}
Coronal Mass ejections (CMEs) are violent phenomena of solar activity with repercussions throughout the entire heliosphere.
Their manifestations into interplanetary space are responsible for major geomagnetic storms, hence the prediction of the arrival of interplanetary coronal mass ejections (ICMEs) at 1AU is one of the primary subjects of the space-weather forecasting \cite[e.g.:][]{Daglis2001, Schrijver+2010}.\\
Several forecasting methods have been proposed over the last two decades.
On one hand, there are the approaches relying on statistical-empirical relations established through observations between coronagraphically measured parameters and quantities related to their heliospheric propagation \citep[e.g. ][]{Brueckner+1998}.
Another approach is represented by numerical MHD-based models of the heliospheric propagation of ICME, generally requiring a detailed knowledge of the state of the heliosphere and large computational facilities \citep[as is the case for the WSA-ENLIL model][]{Enlil99, Enlil04, Enlil1, Enlil2}.
The numerical models are fairly accurate \citep{Vrsnak+2014}, and highly sensitive to the quality of the input parameters \citep{Falk+2010}, as one may expect. 
Recently, the WSA-ENLIL model started to be employed also in a probabilistic approach \citep{Mays+2015, Cash+2015, Pizzo+2015} to quantify the prediction uncertainties and to determine the forecast confidence.
However this approach, interesting as it is, is not widely enough used for real-time space-weather forecasting due to the demanding computational needs.\\
The last category, somewhat lying in between the previous ones, is that employing an MHD- or HD-based simplified description of the interactions an ICME may be subjected to during its interplanetary travel \cite{Gopalswamy+2000, Vrsnak+2002, Michalek+2002, Schwenn+2005}.
Such an approach leads to analytical models or empirical models which require modest computational power.
Those models assume a morphology (either simple as in \citet{Mostl+2011} or more complex as in \citet{Isavnin2016}), a fixed direction and a velocity evolution for the CME and can predict an arrival time and speed from relatively limited initial information on the CME onset conditions.
Such initial conditions can be obtained from several sources, such as LASCO C2 and C3 coronal imagers on-board the \textit{SOlar and Heliospheric Observatory} \citep[SOHO][]{Domingo+1995} and, more recently, from COR1 and COR2 and the \textit{Heliospheric Imager} (HI) on-board the \textit{Solar Terrestrial Relation Observatory} \citep[STEREO][]{Kaiser+2008} spacecraft either separately or using appropriate tools to merge the measures taken from the different instruments and the different points of view \citep[][]{Lugaz+2009, Davies+2012, Mostl+2013, Mostl+2014}.
%\subsection{DBM intro}
In  this paper, we focus our attention on a model belonging to the last category, the \emph{drag-based model} \cite[DBM ][]{Vrsnak+2013}.
The model hypothesizes a simple interaction between the ICME plasma and the solar wind that works to equalize the ICME velocity to that of the solar wind itself.
This is consistent with the measures of the ICME speeds in the near Earth environment which are typically confined in the 400-700km/s range and the estimates of the initial velocity of the plasma ejecta near the Sun, which range between 100km/s and 2000km/s.
This process has been modeled analogously as an aerodynamic or viscous drag by several authors \citep{Cargill+1995, Vrsnak+2002,Shi+2015}.
It makes use of the initial CME velocity, its distance from the Sun at the moment of the measure, and the solar wind speed to compute the travel time at 1AU.\\
%\textbf{THE DBM FAMILY\\}
The DBM has already generated a whole family of approaches, which may differ for the way to evaluate the initial parameters, or how the CME is propagated in the heliosphere.\\
The difference may arise from the peculiarity of the data used (type and source) and from the interpretation of such data to estimate the CME onset parameters, which ultimately depends on the shape assumed for the CME itself (the Fixed-$\phi$ method in \citet{Sheeley+1999, Rouillard+2008,Mostl+2014}, the Harmonic mean method in \citet{Howard+2009,Lugaz+2009}, the Self Similar Expansion method in \citet{Davies+2012,Mostl+2013}, the Graduated Cylindrical Shell method in \citet{Thernisien+2006,Thernisien+2009}, the Elliptic Conversion method in \citet{Rollett+2016}, to cite the most used).
Or, the difference can be in the way the drag effect is approximated and the velocity of the CME evolves as in \citet{Hess+2015}, \citet{Zic+2015} and \citet{Rollett+2016}.\\
In the cited literature, much attention is paid to get the best estimate of the DBM parameters and an evaluation of the associated errors, but none of the mentioned DBM approaches takes into consideration this last information in the implementation of the forecast.\\
%\subsection{Resume}
In this work, we apply a statistical approach on the DBM for the computation of ICME travel times, by introducing the probability distributions rather than exact values for the input parameters.
This approach has the non-trivial advantage to provide also an evaluation of the uncertainty on the arrival time.
In Section \ref{DBM} we rapidly revisit the equations of the DBM model and introduce its probabilistic version.
We also present and discuss the Probability Distribution Functions (PDFs) that we assume to compute the most probable ICME travel times.
Section \ref{data} presents the dataset of CME speeds, onset times and travel times that we use to compute and then compare (Section \ref{test}) the forecast travel times and associated errors.
In Section \ref{conclusions} we comment on our results, comparing with those already present in the literature, and discuss possible applications and further evolutions of this model.\\
For the sake of clarity, we specify that in this work the terms CME and ICME are referred to the plasma and magnetic field structure expelled from the Sun, without the shock that precedes it.\\
\section{The drag-based model} 
\label{DBM}
\subsection{General description}
The drag-based model relies on the hypothesis that all the interactions responsible for the launch of the CME cease in the upper corona, and that, beyond a certain distance, the dynamics of ICME propagation are governed mainly by its interaction with the ambient solar wind.
The DBM considers such an interaction by means of a drag force analogous to that experienced by a body immersed in a fluid.
The idea of an MHD analogous "hydrodynamical" drag is supported by the observation that ICMEs which are faster than the solar wind are decelerated, whereas those slower than the solar wind are accelerated by the ambient flow \citep{Gopalswamy+2000,Manoharan2006}.\\
Following \citet{Cargill2004}, we consider the relative speed dependence of the drag force in the radial direction:

\begin{equation}
F_d = -C_d A \rho(v-w)|v-w|
\end{equation}

where $v$ is the ICME radial speed and ${w}$ that of the solar wind, $A$ is the ICME cross-section, $\rho$ is the solar wind density and $C_d$ is an dimensionless coefficient for the drag force.
In a classical Newton's law framework, this leads to a radial drag acceleration in the form:

\begin{equation}
\label{dragaccel}
a = -\gamma({v}-{w})|{v}-{w}|
\end{equation}

where $\gamma$ is the so-called \emph{drag parameter} which contains the information about the ICME shape, mass, and in general about the effectiveness of the drag effect.\\
Considering the solar wind speed and the drag parameter as constants \citep[which is a good approximation beyond $20-40\,R_{\sun}$ ][]{Cargill2004, Vrsnak+2013}, equation (\ref{dragaccel}) can be solved explicitly, obtaining as functions of time the ICME speed:

\begin{equation}
  	v(t)=\frac{v_0-w}{1 \pm \gamma(v_0-w)t} + w
	\label{vDBM}
\end{equation} 

and the heliospheric distance:

\begin{equation}
r(t)=\pm \frac{1}{\gamma} \ln{\Big[ 1 \pm \gamma(v_0-w)t\Big]} + wt + r_0    
\label{rDBM}
\end{equation}

where the $\pm$ signs apply to the cases $v_0>w$ and $v_0<w$, respectively, and $r_0$ and $v_0$ are the CME distance from the Sun and velocity at the onset time $t_0$. %
In this framework, the model needs four quantities, $[r_0, v_0, \textit{w}, \gamma]$, to compute the heliospheric distance and velocity of the ICME at any $t$.\\
The shape of the ICME we are modeling corresponds to type A) in Fig.9 of \cite{Schwenn+2005}, i.e. the front of the CME is a section of a sphere concentric with the Sun.
\subsection{The Probabilistic drag-based model}
As just stated, the DBM needs four quantities to be computed, namely $[r_0, v_0, \textit{w}, \gamma]$.
The first two quantities suffer from measure errors, while the last two are, in general, unknown.\\
If we consider the measure errors to be described by Gaussian PDFs, and assume a priori PDFs for both \textit{w} and $\gamma$, we can extend the DBM into a probabilistic approach.\\
The Probabilistic drag-based model (P-DBM henceforth), is a Monte-Carlo evaluation of the time of arrival and the velocity of the ICME at a chosen distance from the Sun, transforming the PDFs associated to the inputs into PDFs for the outputs, thus generating best estimates and errors for both the time of arrival and the velocity.
For each ICME whose $r_0$ and $v_0$ are measured, we can generate $N$ different $[r_0, v_0, \textit{w}, \gamma]$ initial conditions sets, randomly chosen from the relative PDFs, to compute via eqs. (\ref{vDBM}) and (\ref{rDBM}) the transit time and the velocity at 1 AU, for example. 
This process generates the PDFs associated to $t_{1AU}$ and $v_{1AU}$, which can be used to estimate the ICME most probable time of arrival and velocity and their associated uncertainties at 1 AU.\\
Of course, the robustness of the results strongly depends on the validity of the assumptions, the realism of the PDFs, and on a thorough exploration of the parameter space, i.e. how large is $N$.
Given the simplicity of eq. (\ref{vDBM}) and (\ref{rDBM}) and the present computing capabilities, $N$ of the order of $10^4 - 10^6$ can be used to explore the parameter space and obtain nicely sampled output PDFs in a matter of seconds. 
\subsection{PDFs for the input quantities}

In this section we introduce the PDFs which will be used for the four input quantities.
\begin{figure*}
 \includegraphics[width=8.8cm]{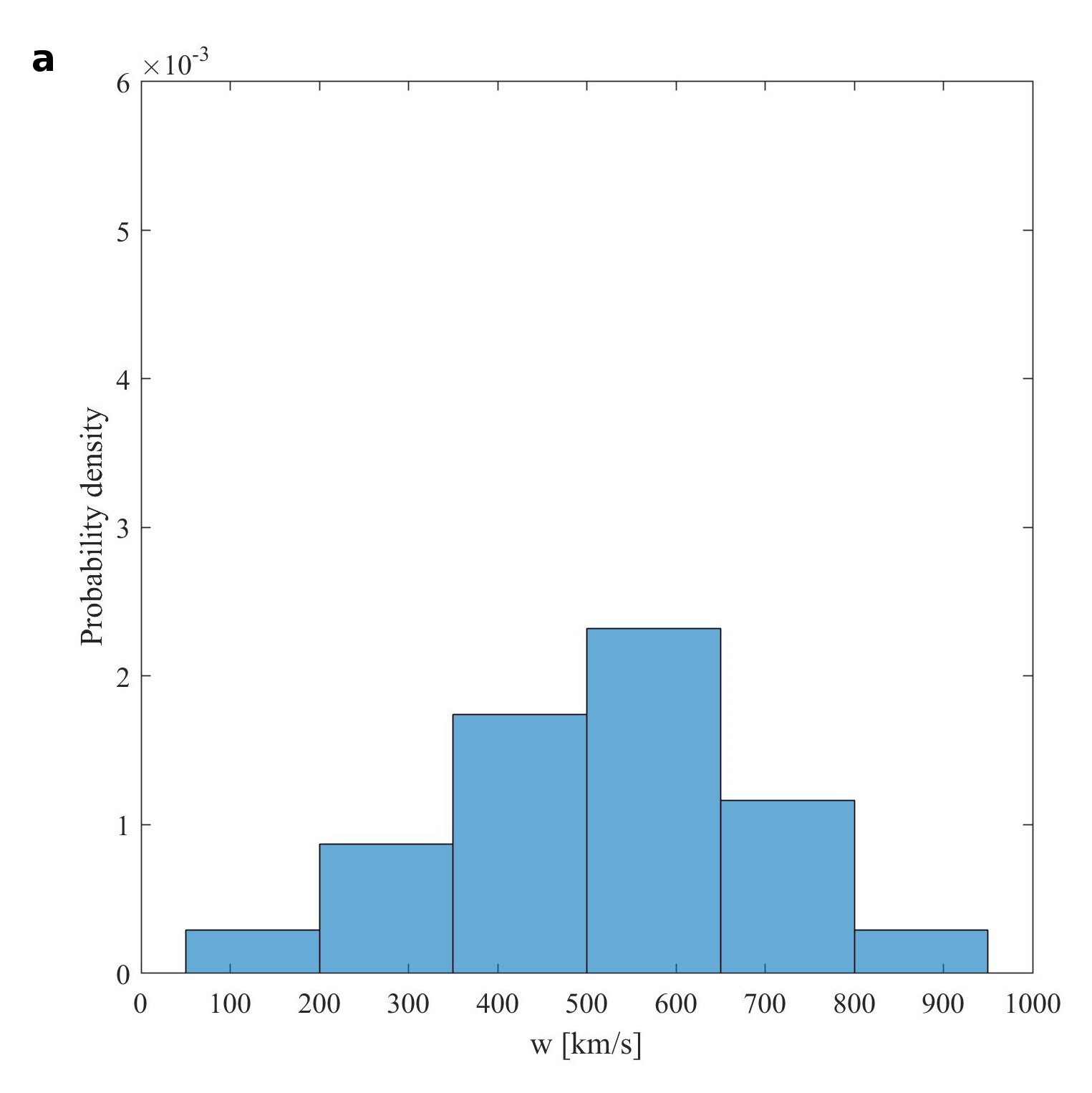}
  \includegraphics[width=8.8cm]{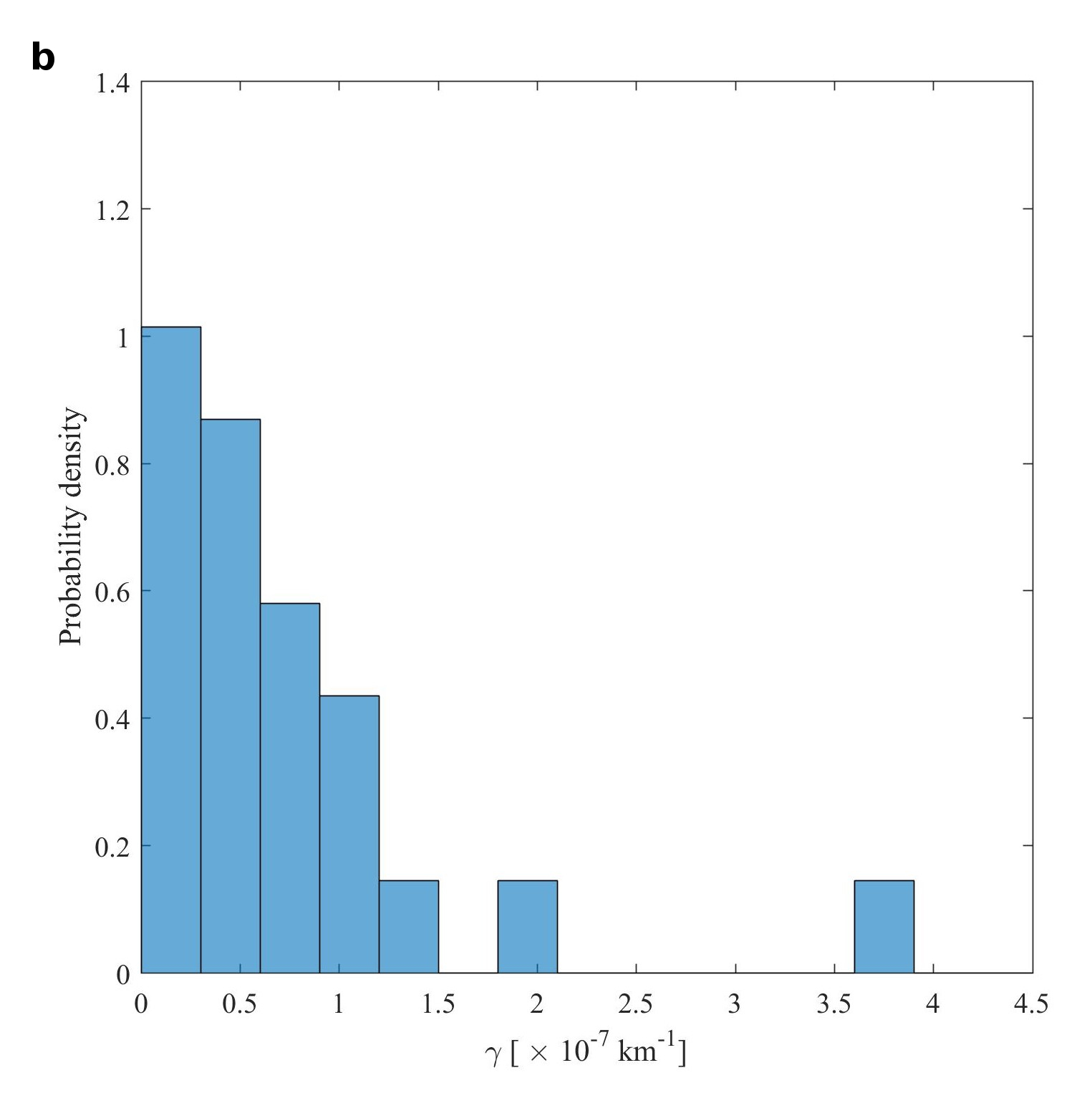}%{gammahist.png}
 \caption{Histograms of $w$ (a) and $\gamma$ (b) obtained by the inversion of \citet{Schwenn+2005} and \citet{Manoharan2006} catalogs.}
 \label{hist_sw}
\end{figure*}
\begin{figure*}
 \includegraphics[width=8.8cm]{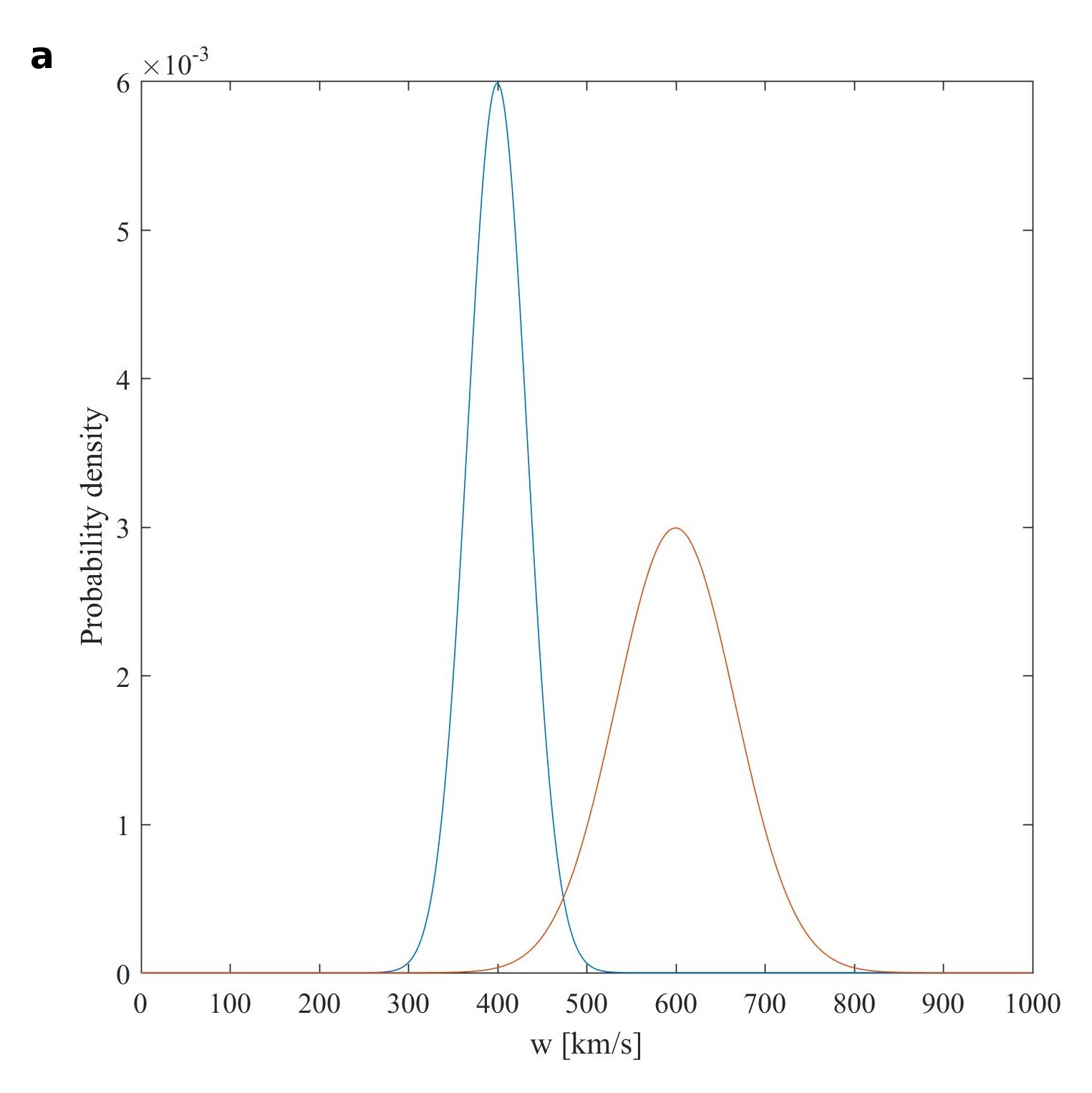}
  \includegraphics[width=8.8cm]{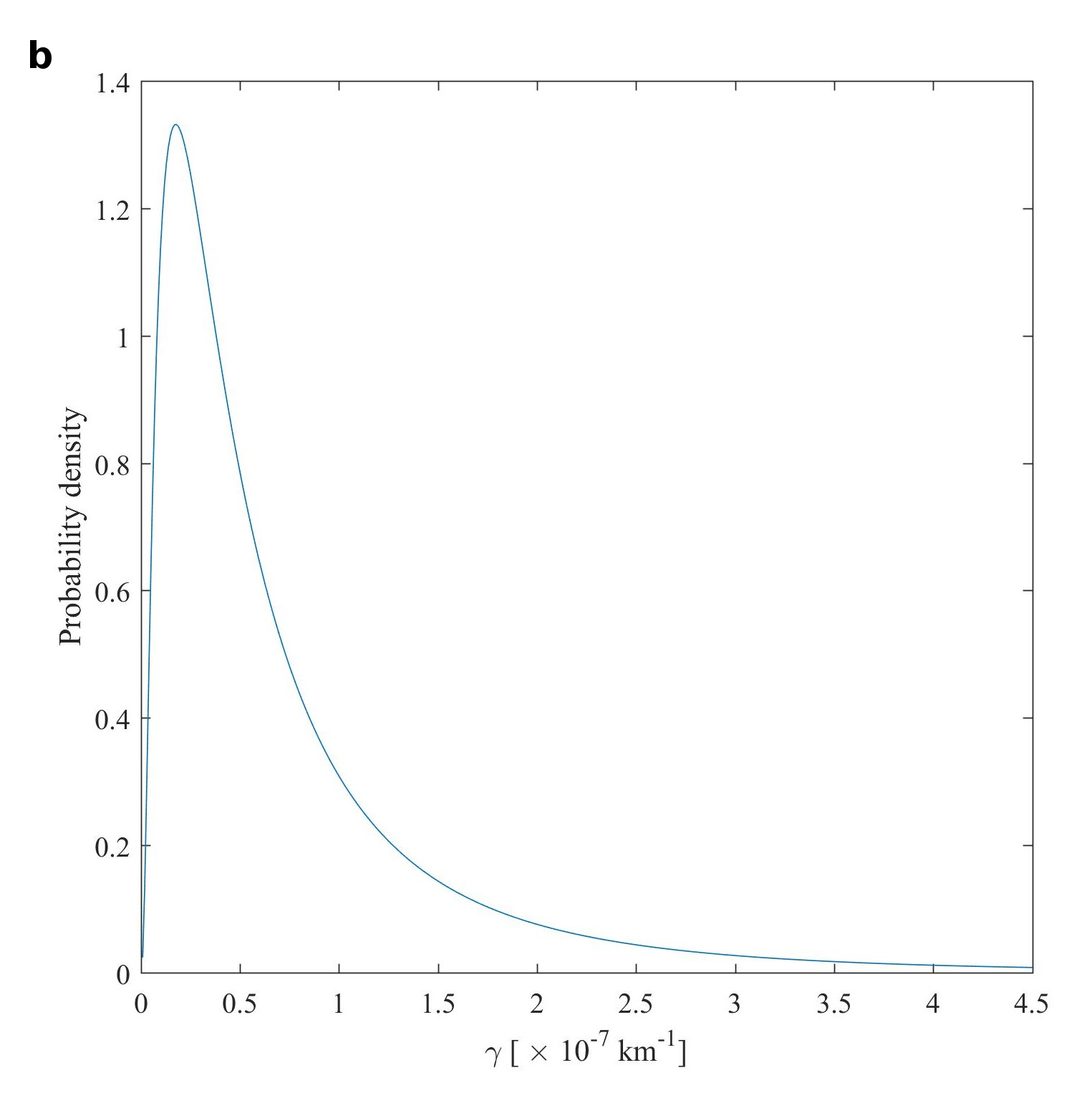}
 \caption{(a) PDF adopted for the random generation of $w$ in the P-DBM, with the slow \textit{w} represented by a Gaussian PDF centered at 400km/s with $\sigma=33~\textrm{km/s}$, and the fast \textit{w} represented by a Gaussian PDF centered at 600km/s with $\sigma=66~\textrm{km/s}$. (b) PDF adopted for the random generation of $\gamma$ in the P-DBM, modeled by a Log-Normal function with $\mu=-0.70$ and $\sigma=1.01$.}
 \label{pdf_sw}
\end{figure*} 
As \citet{Vrsnak+2013} have shown, the two equations (\ref{vDBM}) and (\ref{rDBM}) can be inverted to obtain the drag parameter $\gamma$ and the solar wind speed \textit{w}, if the initial position $r_0$ and speed $v_0$ of an ICME and its time of arrival $t_{1AU}$ and velocity $v_{1AU}$ at 1 AU are known.

%**********************************
\begin{equation}
  	\gamma=\frac{ (v_0-v_{1AU}) } {(v_0 - w) (v_{1AU}-w)t_{1AU}}
	\label{gamma_inv}
\end{equation}
%**********************************

This equation can be used to compute directly $\gamma$ once one has numerically solved:

%**********************************
\begin{equation}
%\frac{1}{\gamma} \ln{[\frac{(v_0-v_{1AU})(v_0-w)} {(v_{1AU}-w)(v_0+w)} +1]} + wt_{1AU} + r_0 - r_{1AU} = 0   
\frac{(v_0 - w) (v_{1AU}-w)t_{1AU}}{(v_0-v_{1AU})} \ln{\left[\frac{(v_0-v_{1AU})} {(v_{1AU}-w)} +1\right]} + w\,t_{1AU} + r_0 - r_{1AU} = 0 
\label{rDBM_inv}
\end{equation}
%**********************************

to obtain $w$.\\
As in \citet{Vrsnak+2013}, we use the catalogs of ICMEs by \citet{Schwenn+2005} and \citet{Manoharan2006} to compute this inversion.
The first list consists of 91 CMEs between 1991 and 2001 for which the authors were able to uniquely associate ICME signatures in front of the Earth after careful inspections of the SOHO/LASCO CME catalog %(https://cdaw.gsfc.nasa.gov/CME_list/)  
and the complete LASCO/EIT data set.
The second list by Manoharan consists of 30 CME events between 1998 and 2004 whose heliospheric evolution has been investigated between the Sun and the Earth using LASCO coronagraphic images and interplanetary scintillation images of the inner heliosphere.
Therefore, these lists include CME events for which a safe association between a remote coronagraphical observation and an in-situ signature has been established, allowing the knowledge of quantities such as transit time and initial and final speed, required for the inversion. 
From the results, we obtain the histograms reported in Fig.\ref{hist_sw} for \textit{w} and $\gamma$.
We choose larger bins than in the original work to make the obtained distributions more robust at the expense of sampling. 
Apart from those differences, these distributions are of course consistent with \citet{Vrsnak+2013} results.\\
For \textit{w}, we can complement the distribution obtained by using the values of the solar wind recorded by SOHO, ACE, ULYSSES, HELIOS \citep{SOHO1, SOHO2, ACE1, ULYSSES1,ULYSSES2, HELIOS1} and many other missions.\\
The common understanding \citep[see][for a review]{Schwenn2006} is that there exist two different PDFs for the so-called slow (below $500~\textrm{km/s}$) and fast solar wind, the latter originated from the coronal holes, which are regions on the Sun with depressed UV emission and low magnetic activity.
The \textit{w} probability densities that we assume are plotted in Fig.\ref{pdf_sw}a, with the slow \textit{w} represented by a Gaussian PDF centered at $400~\textrm{km/s}$ with $\sigma=33~\textrm{km/s}$, and the fast \textit{w} represented by a Gaussian PDF centered at $600~\textrm{km/s}$ with $\sigma=66~\textrm{km/s}$.
Of course, such PDFs are limited to positive values of \textit{w}.
Following the works of \citet{Robbins+2006, Vrsnak+2007}, we adopt the fast \textit{w} PDF in those cases where there is a prominent coronal hole in the center on the disk, the slow \textit{w} PDF in all the other cases.\\
\hspace{1cm} For $\gamma$, we note that the the distribution retrieved by the inversion has a peak in the first bin ($0.2-0.4 \times 10^{-7}$ km$^{-1}$) and then decays, with an extended tail up to $\simeq 4 \times 10^{-7}$km$^{-1}$.
The skewed shape of the distribution suggested to fit ($\tilde{\chi}^2 = 1.13$) such a distribution with the 2-parameter Log-Normal function:

\begin{equation}
\label{Log-Normal}
f(x) = \frac{1}{\sigma\sqrt{2\pi}} e^{- \frac{(\ln x - \mu)^2}{2\sigma^2} } 
\end{equation}

retrieving $\mu=-0.70$ and $\sigma=1.01$, to obtain an analytic form for the PDF, which is shown in Fig.\ref{pdf_sw}b.\\
Despite the fact that we are not putting forward any physical model for the CME kinematics, we must note that the Log-Normal distribution has been found to describe several aspect of solar wind plasma \cite[see][and references therein]{Burlaga+2000} and even the CME speed distribution \citep{Yurchyshyn+2005}. 
In our case, the Log-Normal distribution just provides a good fit to the observed distribution, capturing its properties in just two parameters.\\
%
%\textbf{r= f(LASCO C2 occulter in Solar radii and CME width)  $\pm$ v0*dt0 initial velocity times duration of liftoff of the CME}
%For $r_0$, we have to consider a few points to derive a shape for the PDF.
%First, we assume the geometry of the model by \cite{Xue+2005} for the CME, therefore the distance of the CME apex from the Sun at the moment when the CME becomes visible in the LASCO-C2 FoV can vary from a minimum of $r_0= 1 R_{\odot}$ (if the CME propagation angle is normal to the Sun-Earth direction), to $r_0= 2 R_{\odot}$ in case of a Halo CME whose width is $120\deg$, to $r_0\simeq 2.7 R_{\odot}$ in case of a Halo CME whose width is $60\deg$, or even larger in case of a very narrow CME.
%Then, we recall that \citet{Michalek+2003} retrieved the width distribution for 72 CME from the LASCO catalog, finding that the average width was $120\deg$ (almost twice as large than the value obtained by \citet{Yashiro+2002}).
\hspace{1cm} For $r_0$, we consider that CME detection algorithms have inherent uncertainties for the CME location and the moment and duration of the CME liftoff.
From that, we assume that the PDF of $r_0$ can be modeled by a Gaussian PDF whose average is the last height derived by the CME tracking algorithm at the onset time and whose sigma is estimated from the associated error
\citep[$3\sigma\simeq1 R_{\odot}$ in the case of][]{Shi+2015}.\\
\hspace{1cm} Also for $v_0$, we assume a Gaussian PDF whose average value is the velocity measured by the CME tracking algorithm and whose sigma is the uncertainty associated to the measurement.\\
%
%
% correzione referaggio: riassunto delle operazioni da fare per il tool

\subsection{P-DBM step-by-step}
To resume, here is a step-by-step description of how the P-DBM performs a prediction on the arrival of an ICME:
\begin{enumerate}
\item the position PDF is generated using the last measured CME height within coronagraph images and its associated error;
\item the velocity PDF is generated using the measured velocity and its associated error;
\item the Log-Normal PDF described by Eq.\ref{Log-Normal} ($\mu=-0.70$ and $\sigma=1.01$) is considered for the drag parameter;
\item a Gaussian PDF is chosen for the solar wind velocity, selecting either fast solar wind conditions (600 $\pm$ 66 $km/s$) in the case of a coronal hole in a relevant position of the solar disk, or slow solar wind otherwise (400 $\pm$ 33 $km/s$);
\item N initial condition sets $[r_0,v_0,\gamma,w]$ are randomly generated from those PDFs;
\item N different time of arrivals at 1AU $t_{1AU}$ are computed from Eq. \ref{rDBM}, by setting $t=t_{1AU}$ and $r(t_{1AU})=1AU$, and computing $t=t_{1AU}$ as the root of the equation via an iterative algorithm;
\item the time of arrival PDF is evaluated from the N $t_{1AU}$ values;
\item the best estimate for $t_C$ and its associated error are evaluated as the mean and the root mean square of the time of arrival PDF;
\item Steps 6, 7, 8 are also applied to Eq. \ref{vDBM} to evaluate the best estimate for $v_C$ and its associated error.
%\item values for $r_0$ and $v_0$ are randomly chosen from their respective distributions known from measurements;
% \item a value for $r_0$ is randomly chosen from its distribution known from measurements;
% \item similarly, a value for $v_0$ is randomly chosen from its distribution known from measurements;
% \item a value for $\gamma$ is randomly chosen from the Log-Normal distribution \eqref{Log-Normal};
% \item a value for $w$ is randomly chosen from the solar wind-Fast or -Slow normal distribution, chosen after checking for the presence of coronal hole at the center of the disk;
% \item the single set of values $[r_0,v_0,\gamma,w]$ generated in this way allows the evaluation of the travel time $t_C$ of the ICME at 1 AU through the DBM;
% \item steps 1. - 5. are repeated N times, generating a distribution for the computed travel time (Left panel in Figure \ref{Final}) from which best estimate for $t_C$ and associated error can be evaluated (each point with error bar in right panel of Figure \ref{Final}). 
\end{enumerate}

\section{The Dataset}
\label{data}
In order to test the P-DBM described in the previous section, we use a sample of events from \citet{Shi+2015}.
For such events, a reconstruction of the ICME shape and speed has been obtained with the Graduated Cylindrical Shell model \cite[GCS - ][]{Thernisien+2006, Thernisien+2009} by means of triangulation of coronagraphic images taken from both STEREO and LASCO.
Following \citet{Shi+2015}, we excluded from the original sample those ICME which probably had interactions with the background magnetic field or other CMEs.
We also excluded entry 11 from the original sample which was most probably not correctly associated with the ICME arrival time \cite[cf. ][]{Mostl+2014}.
%The list of events is reported in Table \ref{TongSTEREO1}.\\
The details about how the CGS model has been used to fit the CME shapes and to determine the CME initial speeds and heights are reported in the original paper of \citet{Shi+2015}.
Here, we only recall that the authors estimated the errors of their detection and tracking procedure and that the CME speeds were evaluated through linear fits of the height versus time curves, and the associated error is the uncertainty of the linear fitting.
This will be used to estimate the width of the Gaussian PDFs associated to the CME position and velocity uncertainties and to update the original onset times of the events in \citet{Shi+2015} which are reported in the second column of Table \ref{TongSTEREO}.\\
These onset times are associated to the first detection in the instrument FOV, that is at 2.5 $R_{\odot}$.
In order to employ the DBM in the proper range of heliospheric distances, we choose to move the onset positions at the last useful detection in the instrument FOV at 15 $R_{\odot}$. 
Consequently, we re-evaluated the onset time for each CME by adding a delay of $12.5R_{\odot}/v_0$, using the velocity $v_0$ (third column in Table \ref{TongSTEREO}) obtained by \citet{Shi+2015} through the linear fits of the CME positions exactly between $2.5R_{\odot}$ and $15R_{\odot}$.
The new onset times are reported in the fourth column of Table \ref{TongSTEREO}.
%%%%%% correzione spiegazione del tempo di onset...
%In order to employ the DBM in the proper range of heliospheric distances, the onset time at 2.5 $R_{\odot}$ (first appearance in COR2 FOV) is shifted of the amount %$\Delta t$ required for the ICME to reach the distance of 15 $R_{\odot}$ using the velocity $v_0$ obtained by \citet{Shi+2015} through the linear fit of the running difference images.
%in the COR2 FOV. 
% oppure dico :The time needed for the ICME to travel from 2.5 to 15 R_0 with speed v_0 is evaluated and used to shift the onset time.
%%%%%
%%%%%
Consequently, for the purpose of this work, we can assume for each event a normal distribution of the height $r_0$ at the new onset time, with mean value $<r_0>=15\,R_{\odot}$ and standard deviation $\sigma_r=0.33\, R_{\sun}$.\\
%Analogously, we assume Gaussian PDFs for the initial speeds, with average values equal to the speeds evaluated through the linear fits of the height versus time curves, and the associated error as the uncertainty of the linear fitting.
Furthermore, it must be observed that for the events from the paper by \citet{Shi+2015} the ICME arrival time is referred to the time of first occurrence of an ICME signature in the near Earth environment, which in most cases is the time of arrival of the fore-shock.
To perform a correct validation of the CME transit time forecast, we want to consider the arrival of the ICME leading edge, instead of that of the shock \citep[see also the discussion in][]{Schwenn+2005, Vrsnak+2014}.
To this purpose, for each event, we checked for the time of arrival of a plasma driven effect (Magnetic clouds or Ejecta), as reported in the GMU CME/ICME list compiled by Phillip Hess and Jie Zhang (\url{http://solar.gmu.edu/heliophysics/index.php/GMU_CME/ICME_List}).
In 10 out of 14 cases, we could correct the arrival times.
Column five of Table \ref{TongSTEREO} reports the arrival date and time of the CME, taking into account this update.\\
%For the sake of visualization, the CME indexes, the new onset times, the speed values and the associated errors, the new arrival times are reported in the first four columns of Table \ref{TongSTEREO2}.\\
The last column of Table \ref{TongSTEREO} reports the condition of the solar wind associated with the CME, obtained by the inspection of suitable coronal images and verified by using data recorded by ACE \citep{ACE1}.
\begin{table*}[ht!]
\centering
\footnotesize
{\renewcommand{\arraystretch}{1.5}
\begin{tabular}{c c c c c c}
\# 	&Onset time @ 2.5 $R_{\odot}$&{v}$_{0}\pm 3\sigma$ [km/s] &Onset time @ 15 $R_{\odot}$ 	&Arrival $@$ 1AU &SW\\ %&Transit time (h) 
1 	&12 Dec 2008, 08:37 	&$363 \pm 23$ &12 Dec 2008, 15:16 	&17 Dec 2008, 02:00	&S\\%			&113.4 	
2 	&03 Apr 2010, 09:54 	&$864 \pm 7$ &03 Apr 2010, 12:42  	&05 Apr 2010, 08:00 &F\\%&46.1	\\ 
3 	&08 Apr 2010, 03:39 	&$512 \pm 34$&08 Apr 2010, 08:22 	&12 Apr 2010, 02:00	&S\\		%&80.3  	\\
4 	&16 Jun 2010, 14:39 	&$222 \pm 2$ &17 Jun 2010, 01:32  	&21 Jun 2010, 08:00 &S\\	%&101.3 	\\
5 	&15 Feb 2011, 02:24 	&$769 \pm 12$&15 Feb 2011, 05:33  	&18 Feb 2011, 03:00 &S\\%&72.6 	\\
6 	&04 Aug 2011, 04:39 	&$1512 \pm 90$&04 Aug 2011, 06:15	&06 Aug 2011, 12:00	&F\\%&38.4   \\		
% 7	&06 Sep 2011, 22:54 	&$678 \pm 13$ 			&09 Sep 2011, 15:00$^s$  	&	??\\%&64.1  	\\ eliminato
7 	&14 Sep 2011, 00:39 	&$505 \pm 5$ &14 Sep 2011, 05:26  	&17 Sep 2011, 10:00	&S\\%&73.3  	\\
%8	&22 Oct 2011, 10:39		&$882 \pm 4$ &22 Oct 2011, 13:23 	&25 Oct 2011, 01:00	&S\\%&55.4	\\
8 	&19 Jan 2012, 14:54 	&$1299 \pm 16$&19 Jan 2012, 16:46  	&23 Jan 2012, 00:00	&S\\%&62.1 	\\
9 	&13 Mar 2012, 17:39 	&$1616 \pm 17$&13 Mar 2012, 19:09  	&15 Mar 2012, 19:00	&F\\%&43.4 	\\
10 	&19 Apr 2012, 15:39 	&$607 \pm 15$ &19 Apr 2012, 19:38	&23 Apr 2012, 12:00 &S\\%&82.8 	\\
11	&12 Jul 2012, 16:39 	&$1224\pm14$ &12 Jul 2012, 18:37  	&15 Jul 2012, 06:00	&F\\%&48.4 	\\
12 	&28 Sep 2012, 00:39 	&$1104 \pm 112$&28 Sep 2012, 02:50 	&01 Oct 2012, 12:00	&S\\%&70.3 	\\
13 	&05 Oct 2012, 03:39 	&$558 \pm 21$&05 Oct 2012, 07:59    &08 Oct 2012, 18:00	&S\\%&73.3  	\\
14 	&27 Oct 2012, 16:54 	&$340 \pm 28$&28 Oct 2012, 00:00    &01 Nov 2012, 00:00	&S\\	%&94.1 		 			 		
\end{tabular}
\caption{Sample of events from \citet{Shi+2015} employed to test the P-DBM.
Columns are in order: CME index number, CME onset date and time (UT) at 2.5 $R_{\odot}$, CME initial speed with associated uncertainty, CME onset date and time (UT) at $15 R_{\odot}$, arrival date and time (UT) of the ICME at 1A, solar wind (Slow/Fast) during the CME propagation.}
\label{TongSTEREO}
}
\end{table*}
%**********************************************************************************
%**********************************************************************************

\section{Validation of the P-DBM}
\label{test}
\begin{figure*}
	\includegraphics[width=8.8cm]{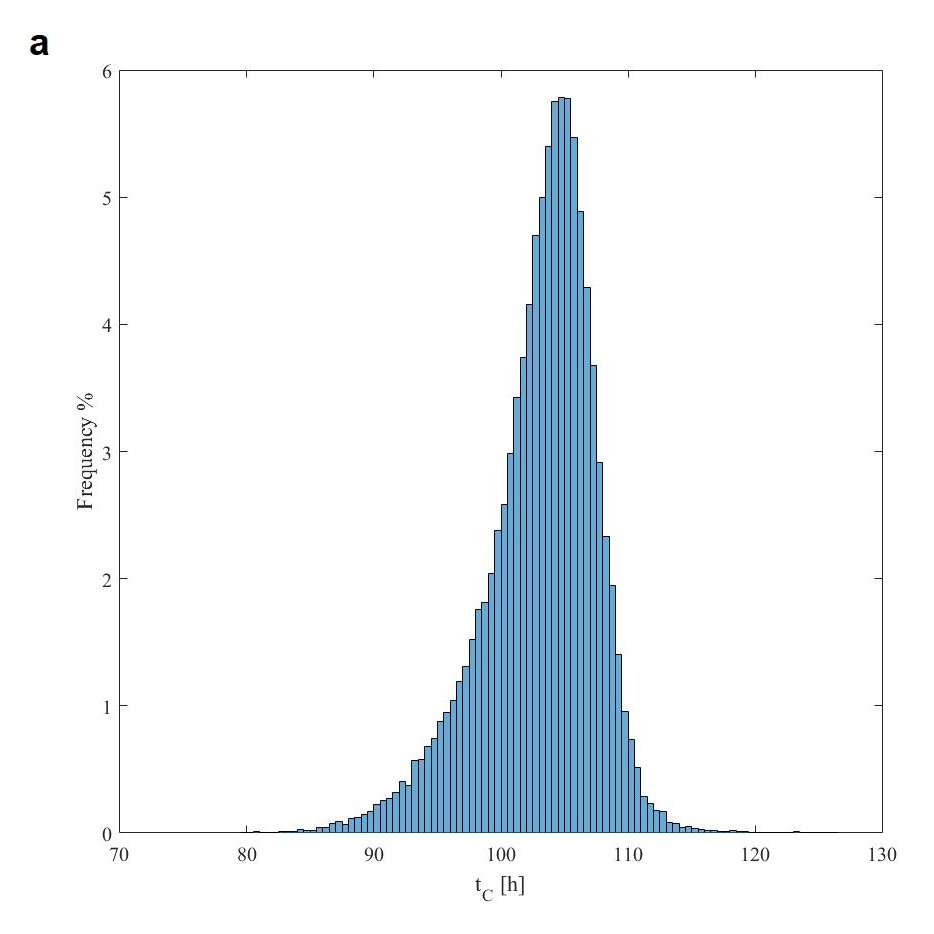}
	\includegraphics[width=8.8cm]{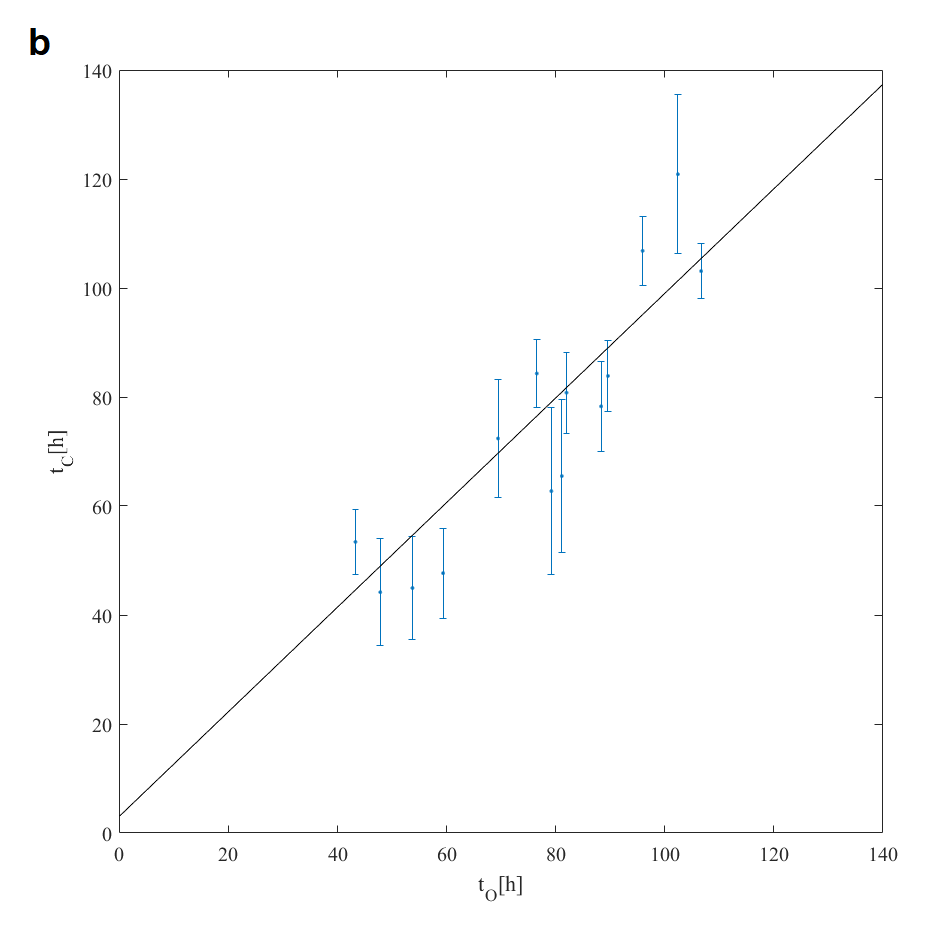}
    \caption{(a) Distribution of the transit times $t_i$ calculated for event \#1 in Table~\ref{TongSTEREO}. N=50000 initial conditions are generated in the P-DBM. (b) Dots with error bars are the forecast transit times $t_C$ versus observed transit times $t_O$. The solid line shows a linear fit to the data.}
   \label{Final}
\end{figure*}

\begin{figure*}
	\includegraphics[width=8.8cm]{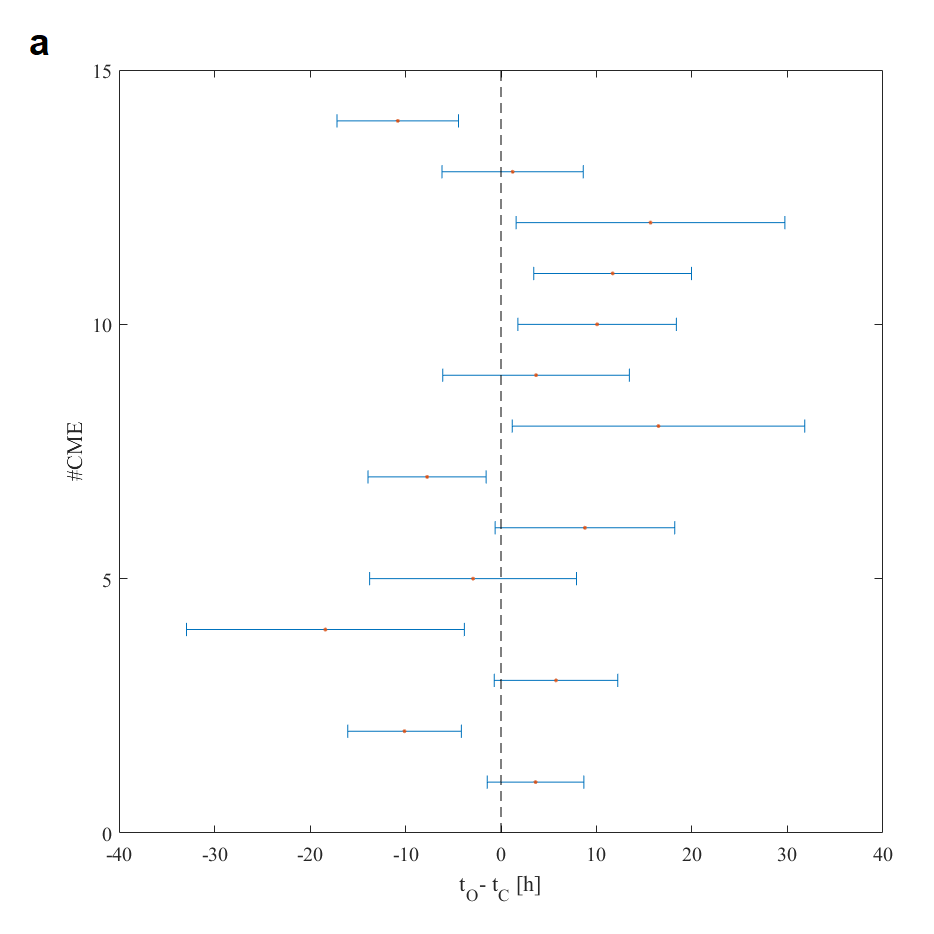}
	\includegraphics[width=8.8cm]{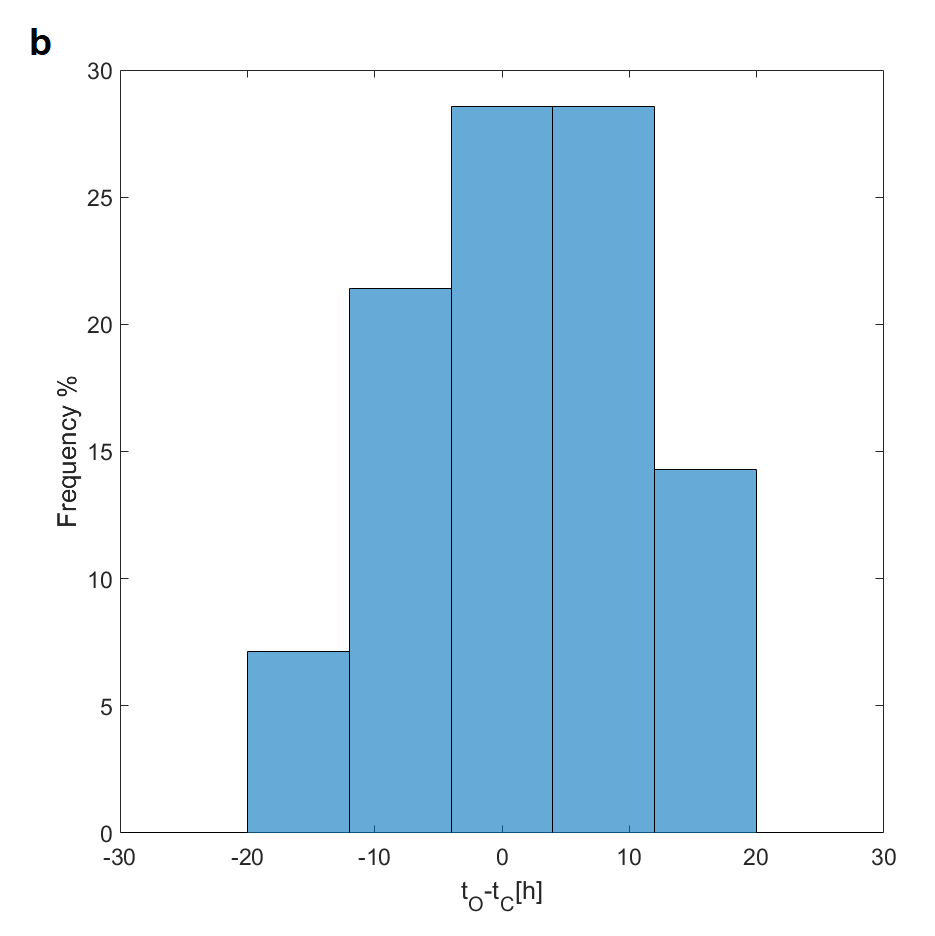}
    \caption{(a) The residuals $t_O-t_C$ and the error associated to $t_C$ for the 14 CMEs. (b) Distribution of the residuals $t_O-t_C$.}
   \label{hist_residui}
\end{figure*}

We apply the probabilistic approach in order to generate the transit time distribution for each event in Table \ref{TongSTEREO}.
For this run, the number of forecast realizations has been set to $N=50000$ and it took less than a minute to obtain the results on a desktop PC.\\
As example, we show in Fig. \ref{Final}a the distribution of the transit times $t_i$ computed by the P-DBM for the first CME of the sample.
As a result from the input distributions, the travel times range from 80h to 120h, with a median value of 103.8h.
The distribution is not symmetric, slightly skewed towards the shorter times.
However, it is viable to describe this distribution by its mean value $t_C=103.1\textrm{h}$ and its root mean square $\sigma=4.4 \textrm{h}$.\\
The results for the whole sample, with $t_C$ and $\sigma$ of the arrival time distributions taken as the measure of the predicted arrival time, are reported in the third column of Table \ref{Table2}.
In the second column, instead, we report the observed transit time $t_O$ computed as the difference between the onset time and the arrival time at 1AU of Table \ref{TongSTEREO}.\\
\begin{table}[ht!]
\centering
\footnotesize
{\renewcommand{\arraystretch}{1.5}
\begin{tabular}{c c c c c }
% Agiornare tabella valori di tO
\# 	&$t_O$ [h]	&$<t_C>\pm \sigma$ [h] & $t_O-<t_C>$ [h] \\
1   &    106.7 &     $103.1\pm  4.4$  &   3.6 \\  
2   &    43.3 &     $53.4\pm  5.8$  &   -10.1 \\  
3   &    89.6 &     $83.9\pm  6.0$  &   5.8 \\  
4   &    102.5 &     $120.9\pm  14.3$  &   -18.4 \\  
5   &    69.4 &     $72.5\pm  10.7$  &   -2.9 \\  
6   &    53.7 &     $44.9\pm  9.3$  &   8.8 \\  
7   &    76.6 &     $84.3\pm  5.8$  &   -7.7 \\  
%8   &    59.6 &     $69.6\pm  12.0$  &   -10.0 \\  
8   &    79.2 &     $62.7\pm  15.2$  &   16.5 \\  
9   &    47.8 &     $44.2\pm  9.7$  &   3.7 \\  
10   &    88.4 &     $78.3\pm  8.1$  &   10.1 \\  
11   &    59.4 &     $47.7\pm  8.1$  &   11.7 \\  
12   &    81.2 &     $65.5\pm  14.0$  &   15.7 \\  
13   &    82.0 &     $80.8\pm  7.0$  &   1.2 \\  
14	&    96.0 &     $106.8\pm  5.8$  &   -10.8 \\  
\end{tabular}
\caption{Results from the P-DBM statistical simulation for the events in Table~\ref{TongSTEREO}. In the first column the CME index as in \ref{TongSTEREO}, in the second column the ICME transit time $t_O$ from 15$ R_{\odot}$ to 1AU, in the third column the computed CME transit time $t_C$ with the associated error $\sigma$. In the fourth column the difference $t_O-t_C$.}
\label{Table2}
}
\end{table}
\noindent
Fig. \ref{Final}b shows a plot of $t_C$ with $1 \sigma$ error bars versus $t_O$ and a least squares linear fit to these data.
The two datasets are evidently highly correlated, with a correlation coefficient $R=$0.87. 
The linear fit performed on the data ($\tilde{\chi}^2=$1.66) retrieved a slope of $1.00\pm0.1$ and a constant value of $3\textrm{h}\pm8\textrm{h}$. 
Given those values, the P-DBM results are compatible with the $t_C=t_O$ hypothesis.\\
Similarly to \citet{Colaninno+2013}, we plot the residuals $t_O-t_C$ and the error associated to $t_C$ for the 14 CMEs in Fig. \ref{hist_residui}a to allow an easy comparison of the forecast results.
In particular, for 7 CMEs out of 14 the forecast residuals are within the error.
Also, we report in Fig. \ref{hist_residui}b the histogram of the residuals.
It can be noted that $80\%$ of the forecasts are within 15h of the actual $t_O$, and just one is beyond 20h.
The distribution is compatible with a Gaussian function, centered in zero and with a $\sigma \simeq 10.6\textrm{h}$, with a marginal partiality towards forecasts behind of the observed times.
To conclude, we computed the average of the absolute value of the residuals $<|\Delta t|>$ = 9.1h, which is often used in the literature to assess the forecast accuracy.\\

\section{Conclusions and future work}
\label{conclusions}
%\subsection{Resume}
In this work, we predicted the transit time between the Sun and the Earth for a sample of 14 CME events.
These events were selected among the database of \cite{Shi+2015}, for which the onset time, initial velocity and transit time are known.
By using the DBM \citep{Vrsnak+2013} and a probabilistic approach, we were able to associate an error to the transit time we computed, assuming that all the input parameters could be described by suitable PDFs.\\ 
For the shape we adopted to model the ICME and since all these ICMEs hit Earth, we did not use the CME principal direction nor the angular width to compute the transit time from the initial parameters.
Nevertheless, it is straightforward to modify the P-DBM to include a different CME shape and to consider the PDFs also for those two input parameters.
Given the very short time needed to compute a CME transit time distribution with this approach, adding two dimensions to the parameter space to be explored should be still feasible with undemanding computational resources.\\
%\subsection{Conclusion}
Even with a model as simple as this, the results of the probabilistic approach are extremely promising:\\
$\cdot$ the scatter plot of $t_C$ vs $t_O$ has a slope which is unity within the errors:\\
$\cdot$ the histogram of the residuals $\Delta~t= t_o -t_C$ has a Gaussian shape, centered in zero and with a $\sigma \simeq 10.6\text{h}$;\\
$\cdot$ the average of the absolute value of the residuals is $<|\Delta~t|>$ = 9.1h.\\
However, less than half of the residuals is within the $1\sigma$  error associated to $t_C$, which is under-performing for a Gaussian distribution of the associated error.
This can either be due to a statistical fluctuation (given the small dimension of the test set) or to an under-estimate of the input PDF widths.
Of course, this disagreement may also arise from the model assumptions. In its simplicity, this DBM implementation models the ICME front as a portion of a sphere concentric with the Sun, therefore neglecting the difference between the ICME apex position and velocity and the ICME position and velocity on the ecliptic plane. On the other hand, this assumption reduces the number of PDFs needed by the model. While it is unclear whether increasing the model complexity will significantly reduce the discrepancy or not, especially considering the intrinsic difficulty in measuring the actual travel times (errors and bias can arise both from the onset and the arrival time estimates), it is instead possible that the PDF we used for $\gamma$, evaluated from actual data, may have incorporated most of such complexity, thus including these effects in the model in a statistical way.
At present, we can conclude that the chosen PDFs led to good estimations of the average times on transit time forecasts, but we need a larger sample to properly evaluate the robustness of the associated errors.
This is probably the main task for future work.
\\
% instead 
There is a vast literature to compare our results with.
We limit ourselves to cases where the authors employed data with projection effects eliminated (measures in quadrature or multi-spacecraft plus CGS model), as in our test.\\
\cite{Gopalswamy+2001} found an empirical relation between the initial CME velocity and its  acceleration and applied this relation to a model to compute the Time of Arrival (ToA) at Earth.
They were able to forecast the ICME ToA at 1AU with a mean error $<|\Delta~t|> = 10.7\textrm{h}$ and $72\%$ of the events had ToA within $\pm 15\textrm{h}$ from the predicted values\\
\cite{Owens+2004} tested on a 35 CME sample three different models: a model with a constant acceleration \cite{Gopalswamy+2000}, a model with an acceleration which ceased before 1AU \cite{Gopalswamy+2001}, and the original aerodynamic drag model \cite{Vrsnak+2002}.
These three model were best fitted on the sample and their $<|\Delta~t|>$  varied from 12h to 9h.\\
\cite{Schwenn+2005} derived an empirical correlation between halo CME expansion speeds and travel times to 1AU, fitting a straight forward deceleration model assuming viscous drag on the data from 75 halo CME events.
For $95\%$ of those events, the shock associated to the CME arrived within $\pm 24\textrm{h}$ of the predicted time.\\ %(but see their statements about the definition of events and ToA).\\
\cite{Colaninno+2013} found that a first-order polynomial to the height-time measurements beyond $50 R_{\sun}$ (0.23 AU) was the best parameter for predicting the CME ToA at 1AU. 
For a sample of 9 CME, they were are able to predict their ToA to within $\pm ̇13\textrm{h}$. It is worth to stress that they supplemented their data with STEREO/HI observations, thus increasing the accuracy of their CME initial parameter estimation.\\
\citet{Taktakishvili+2009} instead evaluated the performances of the ENLIL MHD simulation fed with a cone model of CME for a sample of 14 events. They reported an average absolute error of 6 hours, which is also very similar to the error reported by \citet{Millward+2013} of 7.5 hours, obtained again with ENLIL simulation initialized with CME parameters obtained via the CME Analysis Tool (CAT), but on a larger (25 events) set.
\cite{Vrsnak+2014} compared the CME arrival time prediction based on the DBM against ENLIL.
They reported estimation errors of about 14h with standard deviation ranges from 14h to 19h, depending on the sample and method.\\
\cite{Shi+2015} used a multi-parametric best fit on the transit time versus the initial speed for different drag based regimes.
Depending on the regime, they were able to reach a mean error $<|\Delta~t|>$ down to 6.7h.
Since we employed exactly their sample to test the P-DBM, we can note that their model performed better than ours on this sample.\\
It is worth to stress that \cite{Shi+2015} (and all the authors previously cited) fitted their distribution to the data, therefore optimizing the model to that dataset.
Our approach, in contrast, used two datasets to build the PDFs and was tested against an independent dataset, thus providing a true a-priori forecast test.\\
%Our approach, in contrast, does not fit any parameter to the data, thus providing a true a-priori forecast test.\\
%
%All assume self-similar expansion of the CMEs\\
%All use data with projection effects eliminated (measure in quadrature or multi-spacecraft+CGS model)\\
%Due to the poor statistic obtained from inversions we deliberately excluded eventual correlations between initial conditions and parameters which are to be expected and which in principle could reduce the width and uncertainty on the generated transit time distribution, will be the object of future works.
%
%\subsection{future works}
As already stated, we are aware that our results are based on the analysis and comparison of a very limited dataset.
Among the next steps in the further validation of this approach, is the test with a larger database of ICME.
Since databases which provide information sufficient to fully characterize the ICME are difficult to retrieve, we are already taking into consideration the possibility of having much less information on the ICME onset and morphology.\\
Therefore, we are working to include both the uncertainty on the angular extension, the uncertainty on the main direction and on the de-projected velocity of the CME in the P-DBM, again, modeled by PDFs.
At present, we are also working on a real-time implementation of the P-DBM which ingests the parameters of ICMEs tracked by the CACTUS software \citep{Robbrecht+2004} and forecast the time of arrival at 1AU of the ICMEs and their velocity, of course with the associated errors.
As a result, we will build up a database of the results and we plan to verify and possibly re-consider the PDFs we have chosen for the input parameters.\\
Also, we are pondering an evolution to consider a different morphology of the ICME, passing from the cone model (and its intersection with the ecliptic plane) to a 3-D light-bulb model or a similar model \citep[e.g.:][]{Kleimann2012}.\\
All these effects significantly alter the travel time and it is worth to explore how the P-DBM can include them in its probabilistic approach.\\
Since a complete and real-time stereoscopic determination of the CME morphology and propagation will not be available in the near future, all the 3D effects which are not taken into account in the present P-DBM should be considered as partially unknown variables 
and should be modeled by suitable PDFs, constrained by as much information as available.
As example, the real width, direction and velocity of the CME have to be evaluated from images which suffer from projection effects. There are several ways to de-project the data, which imply different assumptions. One of the simplest \citep{Zhao+2002} assumes that the cone-shaped CME has its vertex in the Sun's center and its axis normal to the solar surface at the position of a relevant solar magnetic feature (erupting filament or flaring AR). In such a case, this information, error propagation theory and previous CME parameters statistics could be used to generate the PDFs needed to propagate the ICME with the P-DBM.
It is likely that adding other uncertainties from other input parameters will enlarge the error associated with the forecast, but it is important to stress again that the P-DBM light computation needs make it interesting to evaluate the propagation of any ICME in any portion of the inner solar system.\\
To conclude, the accurate prediction of the time of arrival of an ICME to Earth or other interesting part of the Heliosphere \citep[e.g.:][]{Falkenberg+2010MARS} is of critical importance for our high-technology society and for any future manned exploration of the solar system.
We think that as critical as the prediction accuracy is the knowledge of precision, i.e. the error associated to the forecast.
The method we presented here, building on the DBM model of \cite{Vrsnak+2013}, is capable to predict the arrival time of ICMEs to the Earth and its uncertainty with minor computation necessities, providing a forecast of the space weather in the near Earth environment with a 2-day horizon.\\
\acknowledgement{This research work has been partly supported by the Italian MIUR-PRIN grant 2012P2HRCR on "The active Sun and its effects on Space and Earth climate" and by Space Weather Italian COmmunity (SWICO) Research Program, from the Regione Lazio FILAS-RU-2014-1028 grant on "Banca Dati di Space Weather da Strumenti nello Spazio ed a Terra", and from the EC Tender No.434/PP/GRO/RCH/15/8381 for the "Ionosphere Prediction Service".\\
%GN acknowledges the PhD School of Universit\`a degli studi dell'Aquila for providing the opportunity to work on this project.\\
GN wishes to take this opportunity to express his sincere appreciation for the PhD grant from the Universit\`a degli Studi dell'Aquila, for the supplies and facilities placed at his disposal, and for providing the opportunity to work on this project.\\
The authors thank R. Schwenn for sharing the CME database used in \cite{Schwenn+2005}.\\
The authors thank the anonymous referees for their insightful and helpful comments on earlier versions of the paper.
The editor thanks two anonymous referees for their assistance in evaluating this paper.\\
}

\bibliography{PDBM}   % bibliography data in PDBM.bib
\bibliographystyle{swsc}
\end{document}